# Structural Color 3D Printing By Shrinking Photonic Crystals

**Authors:** *Yejing Liu,[1] Hao Wang,[1] Jinfa Ho,[3] Ryan C. Ng,[2] Ray J. H. Ng,[1,3] Valerian H. Hall-Chen,[4] Eleen H. H. Koay,[3] Zhaogang Dong,[3] Hailong Liu,[1] Cheng-Wei Qiu,[5] Julia R. Greer,[2] Joel K. W. Yang,[1,3,*]*

[1] Engineering Product Development, Singapore University of Technology and Design, Singapore 487372.

[2] Division of Engineering and Applied Science, California Institute of Technology, Pasadena CA 91125, United States of America.

[3] Nanofabrication Department, Institute of Materials Research and Engineering, Singapore 138634.

[4] Rudolf Peierls Centre for Theoretical Physics, University of Oxford, Oxford OX1 3PU, United Kingdom.

[5] Department of Electrical and Computer Engineering, National University of Singapore, Singapore 117583.

* To whom correspondence should be addressed. Email: joel_yang@sutd.edu.sg

## Abstract

The rings, spots and stripes found on some butterflies, Pachyrhynchus weevils, and many chameleons are notable examples of natural organisms employing photonic crystals to produce colorful patterns. Despite advances in nanotechnology, we still lack the ability to print arbitrary colors and shapes in all three dimensions at this microscopic length scale. Commercial nanoscale 3D printers based on two-photon polymerization are incapable of patterning photonic crystal structures with the requisite ~300 nm lattice constant to achieve photonic stopbands/ bandgaps in the visible spectrum and generate colors. Here, we introduce a means to produce 3D-printed photonic crystals with a 5x reduction in lattice constants (periodicity as small as 280 nm), achieving sub-100-nm features with a full range of colors. The reliability of this process enables us to engineer the bandstructures of woodpile photonic crystals that match experiments, showing that observed colors can be attributed to either slow light modes or stopbands. With these lattice structures as 3D color volumetric elements (voxels), we printed 3D microscopic scale objects, including the first multi-color microscopic model of the *Eiffel Tower* measuring only 39-µm tall with a color pixel size of 1.45 µm. The technology to print 3D structures in color at the microscopic scale promises the direct patterning and integration of spectrally selective devices, such as photonic crystal-based color filters, onto free-form optical elements and curved surfaces.

**Introduction**

Realizing the full potential of 3D photonic crystal structures with wide-ranging applications in integrated optical components,[1-3] 3D photonic integrated circuitry, anti-counterfeiting security labels[4,5] and dye-free structural color printing requires the ability to pattern and position these crystals deterministically. However, this ability to position such structural colors at will has eluded us while certain biological species have evolved to be masters of structural color. The fabrication of such 3D structures still remains a challenge, involving manual stacking of 2D structures with stringent alignment requirements.[6] Additive manufacturing via 3D printing removes the need for this cumbersome assembly process, thus enabling the deterministic fabrication of complex 3D photonic structures. These structures enable control of the optical path, polarization and amplitude of certain wavelengths of light at the sub-microscopic scale, resulting in enhanced or new optical properties that are not observed in naturally occurring materials.[4,7-14]

The lattice constants of photonic structures made by direct laser writing are in the micrometer length scale and operate in the infrared (IR) spectral region. To extend the operational window of these photonic devices to the UV-visible spectral range (100-700 nm), the lattice constants of the photonic crystals must be reduced accordingly, and the constituent material should have as high a refractive index as possible. Structural colors arising from the interaction of light with photonic structures are of great interest as they do not degrade and can be printed at high-resolutions compared to colors from pigments and dyes.[15-19] Unlike colloidal self-assembly approaches[17-24] and 2D full-color prints using electron-beam lithography,[25-28] lithographic direct-laser writing (DLW) allows precise pattern placement in all three dimensions, enabling the production of a continuous hue of 3D structural photonic crystal colors by controlling its lattice constants.

Commercially available DLW printing techniques lack the necessary resolution to fabricate 3D photonic structures with stopbands/ bandgaps in the visible range. For instance, the Nanoscribe GmbH Photonic Professional GT can achieve ~500 nm lateral resolution and 200 nm linewidths with the proprietary IP-Dip resist. This resolution is limited by diffraction, material structural rigidity and accumulation of below-threshold exposure in the photoresist that induces unwanted polymerization in surrounding areas. To improve the resolution of 3D two-photon DLW, stimulated emission depletion (STED) and diffusion-assisted high resolution DLW approaches have been demonstrated, producing gyroid and woodpile photonic crystals with 290 nm and 400 nm lattice constants, respectively.[29-32] Nonetheless, the throughput of these approaches are low, the processes require custom resists and complex systems consisting multiple laser sources with precise beam alignment. While heat-shrinking methods have shown dramatic size reduction,[33,34] none have been suitable for producing structures fine enough to generate structural color. Currently no widely-available 3D printing technique exists that can achieve spatial resolutions better than ~300 nm with a single-wavelength laser beam and reasonable throughput.

Figure 1A introduces a shrinking method to enable the direct 3D printing of photonic crystals with stopbands in the visible range. Our approach combines the printing of 3D photonic structures using the Nanoscribe GmbH Photonic Professional GT and the IP-Dip resist, followed by a heat-induced shrinking process. To demonstrate the capability of this process, we fabricated woodpile photonic crystal structures with lattice constants as small as ~280 nm, a dimension comparable to the finest periodicities in butterfly scales. The refractive index of the cross-linked polymer increases in the process, which is desirable for widening the stopbands for structural color applications. As evidence of successfully fabricated photonic crystals, photonic stopbands appear in the visible range and the woodpile photonic crystal structures reflect vivid colors with hues dependent on their lattice constants. We produced full-color prototypes of the *Eiffel Tower* with voxel sizes as small as 1.45 µm (*x-y*) and 2.83 µm (*z*). To the best of our knowledge, this is the first demonstration of a full-color 3D printed object based on dielectric structural colors instead of dyes. The height of the 3D printed *Eiffel Tower* model was as small as 39 µm, demonstrating the capability to design and fabricate 3D photonic crystals that are shrunk to fit specific colors. This technology would be broadly applicable to achieve compact photonic optical devices and metasurfaces, such as 3D integrated circuits on chips, and photonic polarizers on optical fibers that can precisely control the wavelengths and polarizations of the output light.

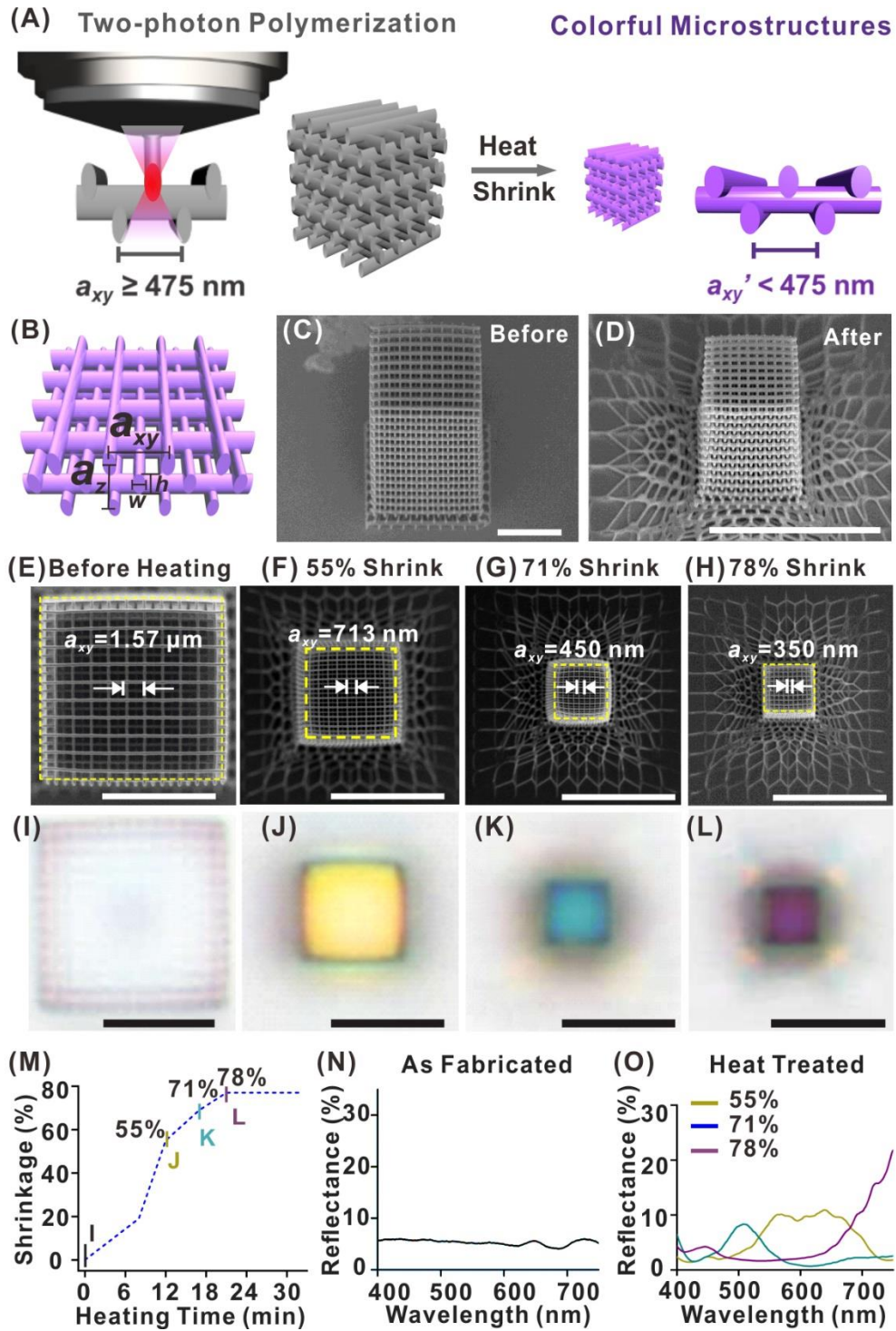

**Figure 1.** Heat-shrinking induced colors of 3D printed woodpile photonic crystals. (A) Schematic of the fabrication process. Left: woodpile photonic crystal written in commercial IP-Dip resist by two-photon polymerization at dimensions well above the resolution limit of the printer to prevent structures from collapsing. Right: after heat treatment, the dimensions of the photonic crystal are reduced below the resolution limit of the printer, and colors are generated. The colors change with different degrees of shrinkage. (B) Schematic showing 1 axial unit of the woodpile structure. $a_{xy}$ and $a_z$ denote the lateral and axial lattice constants, respectively. Tilted-view scanning electron micrographs (SEM) of a representative woodpile photonic crystal (C) before and (D) after heating. SEM images and corresponding bright field reflection-mode optical micrographs of the woodpile photonic crystal before heating (E,I) and with shrinkages of 55% (F,J), 71% (G,K) and 78% (H,L). (M) Shrinkage of the woodpile photonic crystal heated at ~450 °C as a function of heating duration. Reflectance spectra of the woodpile photonic crystal (N) before heating and (O) after heating with 55%, 71% and 78% shrinkage. Scale bars represent 10 µm.

## Results and Discussion

### Heat-induced photonic crystal colors

We first investigate the ability of this process to produce structures at resolutions unachievable solely by conventional DLW. The woodpile lattice structure was chosen as its design is conveniently scripted and patterned rapidly by DLW. It consists of orthogonal grating stacks as shown in Figure 1B, where $a_{xy}$ and $a_z$ denote the lattice constants of the woodpile structure in the lateral and axial directions, while $w$ and $h$ denote the width and height of the constituent rods. Polymeric woodpile photonic crystals with a range of $a_{xy}$ and $a_z$ were directly printed using a two-photon lithography system (Nanoscribe GmbH Photonic Professional GT) and the commercial acrylate-based photoresist IP-Dip™. Structurally stable woodpiles with clearly separated rods were obtained for $a_{xy}$ as small as ~ 475 nm (Figure S1A). Next, the photonic crystals were heated to 450 ± 20 °C in an Ar gas environment where the cross-linked polymers underwent time-dependent decomposition, which reduced the size of the photonic crystals by up to ~80% linear shrinkage (Figure 1). This shrinkage allows us to produce structures with a minimum lattice constant of ~280 nm, clearly beyond the resolution limit of the DLW system. Note that not all of the structures could be successfully shrunk, e.g. structures with lattice constants less than 1.1 µm would simply coalesce into a homogenous particle during the heating process.

In addition to shrinkage, heating also alters the effective shape of the laser writing spot. The laser writing spot at the Gaussian focal point of the DLW system is ellipsoidal, thus resulting in a vertical resolution that is ~3x worse than the lateral resolution, and line structures having an elliptical cross section (see Supporting Figure S2A). The heat-shrinking process effectively produces a more spherical writing spot (see Supporting Figure S2B) allowing for a minimum z-axis resolution of ~380 nm, significantly below the two-photon Sparrow criterion in $z$-direction of ~500 nm. The fabrication reliability and reproducibility for smaller structures is also improved as we can pattern mechanically robust structures within a larger process window (see Figures S2-S8 for additional information on the heat shrinking process). This concept is similar to that demonstrated in 2D with Shrinky Dinks, where structures printed using a simple desktop printer were later heat shrunk to micron length scales.[35] The printed complex gyroid and diamond lattices (Figure S2D and E) demonstrate the generality of the approach.

To determine the amount of shrinkage as a function of heating duration (Figure 1M), we fabricated several structures with identical nominal parameters as shown in Figure 1C and heated them for different durations. The structure has 12 repeat layers (48 stacks) with initial $a_{xy}$ = 1.57 µm, $a_z = \sqrt{2}a_{xy}$, and is comprised of rods with initial $w$ of 330 nm and $h$ of 1.1 µm. Figures 1E-H and I-L are SEMs and corresponding optical micrographs of the structures after 12 min, 17 min, and 21 min of heating at 450 °C, clearly showing a lateral shrinkage of 55% ($a_{xy}$ = 713 nm), 71% ($a_{xy}$ = 450 nm) and 78% ($a_{xy}$ = 350 nm) respectively. The rod width $w$ decreased to ~100 nm after 21 min of heating (Figure 1H). After heat shrinkage, the bottom three repeat units turned into an intricate net-like mesh due to the adhesion of the bottom-most layer to the substrate (Figure 1D).

However, the top nine repeating units were observed to be uniform after shrinking as they are sufficiently far from the substrate. The three bottom units are analogous to "rafts" in fused deposition modeling (FDM) 3D printing used to improve adhesion to the print bed, and the top nine units as a uniform photonic crystal in subsequent analyses. Alternatively, uniform structures can be achieved by printing a thick solid block underneath the structure as a buffer layer to decouple the strain mismatch between the substrate from the structure (Figure S6).

Next, we investigate the evolution of cross-linked IP-Dip with the heating process and propose a mechanism for the shrinkage. From thermogravimetric analysis (Figure S7B), we observed that the largest reduction in mass occurred at 450 °C, thus all our samples were heated to this temperature. Raman spectroscopy measurements (Figure S7C) of IP-Dip before heating show the characteristic $CH=CH_2$ stretching mode at 1632 cm$^{-1}$. This peak corresponds to unreacted terminal alkene groups, indicating the presence of partially cross-linked polymers or unreacted monomers in the photonic crystal structure. After heating, the peak disappears, suggesting that these components were removed upon heating. We further observe that heating reduces the peak intensities of the C=O (1722 cm$^{-1}$), C-O (935 cm$^{-1}$) and C-H (2947 cm$^{-1}$) stretching modes, while introducing peaks corresponding to activated (porous) carbon at 1593 cm$^{-1}$ (graphitic carbon), 1353 cm$^{-1}$ (disordered carbon) and 2500-3100 cm$^{-1}$ (sp$^2$ rich carbon).[36,37] These observations indicate that the IP-Dip polymer is at the onset of carbonization, and that carbon oxide, water vapor, and small molecules of unlinked monomers are released from IP-Dip during heating. This results in structures with decreased volume and increased density and carbon content. Unlike glassy carbon that forms at higher temperatures in larger structures, the presence of the C=O and C-O bands in the material after heating indicates that the material was not entirely converted into solid carbon at this temperature. Ellipsometric measurements of heat treated IP-Dip film show an increase in the refractive index ($n$) from 1.59 to 1.82 (at 400 nm, Figure S7D) accompanied by an increase in the extinction coefficient ($k$) from ~ 0 to 0.2 (at 400 nm, Figure S7E), further corroborating the increase in density and carbon content in the photonic crystal structure. Compared to previous works that pyrolyzed IP-Dip at 900 °C,[34] our heat treatment process achieved an almost identical amount of volume shrinkage at a much lower temperature of 450 °C. The low temperature prevents IP-Dip from turning into glassy carbon with high optical losses (with $k = 0.8$ at 400 nm), and allows us to maintain a relatively low $k$ while increasing $n$, which is desirable for making photonic stopbands/ bandgaps. Attempts to pyrolyze woodpile structures at 900 °C resulted in the complete decomposition of the woodpile (Figure S8). Despite the large dimensions used, ($a_{xy} = 1.9$ µm and $w = 403$ nm) the rods could have been too thin or insufficiently cross-linked to survive the process. Heating at 900 °C therefore imposes additional limitations on the minimum rod width in order to maintain structural integrity.[37]

The reduction in lattice constant and increase in $n$ of the photonic crystal resulted in colors that evolve with the degree of shrinkage. Before heating, the reflectance of the woodpile photonic crystal is weak and no colors were observed (Figures 1I, N). Colors emerge as the structures shrink, shifting from yellow to blue and purple for linear shrinkage values of 55%, 71% and 78% (Figures 1J-L). The colors observed from the reflection mode optical micrographs agree with the measured spectra (Figure 1O), with the reflection peak center shifting from ~ 600 nm (55% shrinkage) to ~ 508 nm (71% shrinkage) and ~ 445 nm (78% shrinkage). At 78% shrinkage, an additional strong reflection peak was observed at ~ 780 nm.

The bandstructure calculations of the woodpile photonic crystals provide insight into the reflectance spectra and observed color (Figures S9 and S10). Without loss of generality, we consider only the bandstructures in the Γ-X and Γ-K directions, corresponding to top-down and side illumination, respectively. For the woodpile photonic crystal with $a_{xy}$ = 1.57 μm (before heating), the large number of photonic states form a continuum in the visible-IR range (Figure S9A) and visible light can propagate through the photonic crystal, resulting in low reflectance and a colorless appearance. After heat shrinkage to $a_{xy}$ = 350 nm, the woodpile photonic crystal exhibits angle-dependent colors. As shown in the optical micrographs in Fig. 2A, the cubic structure appears maroon from the top facet but yellow from the side. The corresponding reflectance spectra are shown in Fig. 2B with brightfield imaging configurations inset. Along the Γ-X direction of the bandstructure (Fig 2C), a stopband is present at ~750 nm near infra-red (NIR) region, corresponding to the strong reflection peak at ~780 nm observed experimentally. In addition, several states with inflection points (i.e. $\frac{d\omega}{dk} = 0$) indicative of slow light modes are present at ~430 nm. Due to impedance mismatch between the incident light and these slow light channels, coupling to these modes is poor,[38] resulting in the reflection peak measured at ~450 nm. The structure thus appears maroon under top-down illumination due to the combination of the slow light reflection peak at ~ 450 nm (blue) and a small spectral contribution from the tail of the strong NIR stopband reflection around 750 nm (red). The stopband along Γ-K is blueshifted relative to the stopband along Γ-X (750 nm → 680 nm) in agreement with the shift in measured reflectance peaks from the side (~75° tilt) relative to normal incidence (780 nm → 740 nm). Slow light modes were also present in the 400-450 nm, 550-650 nm, and 700-725 nm regions. Spectral features were observed at 550-650 nm in both reflectance measurements and calculated bandstructures, producing the yellow color observed under side illumination.

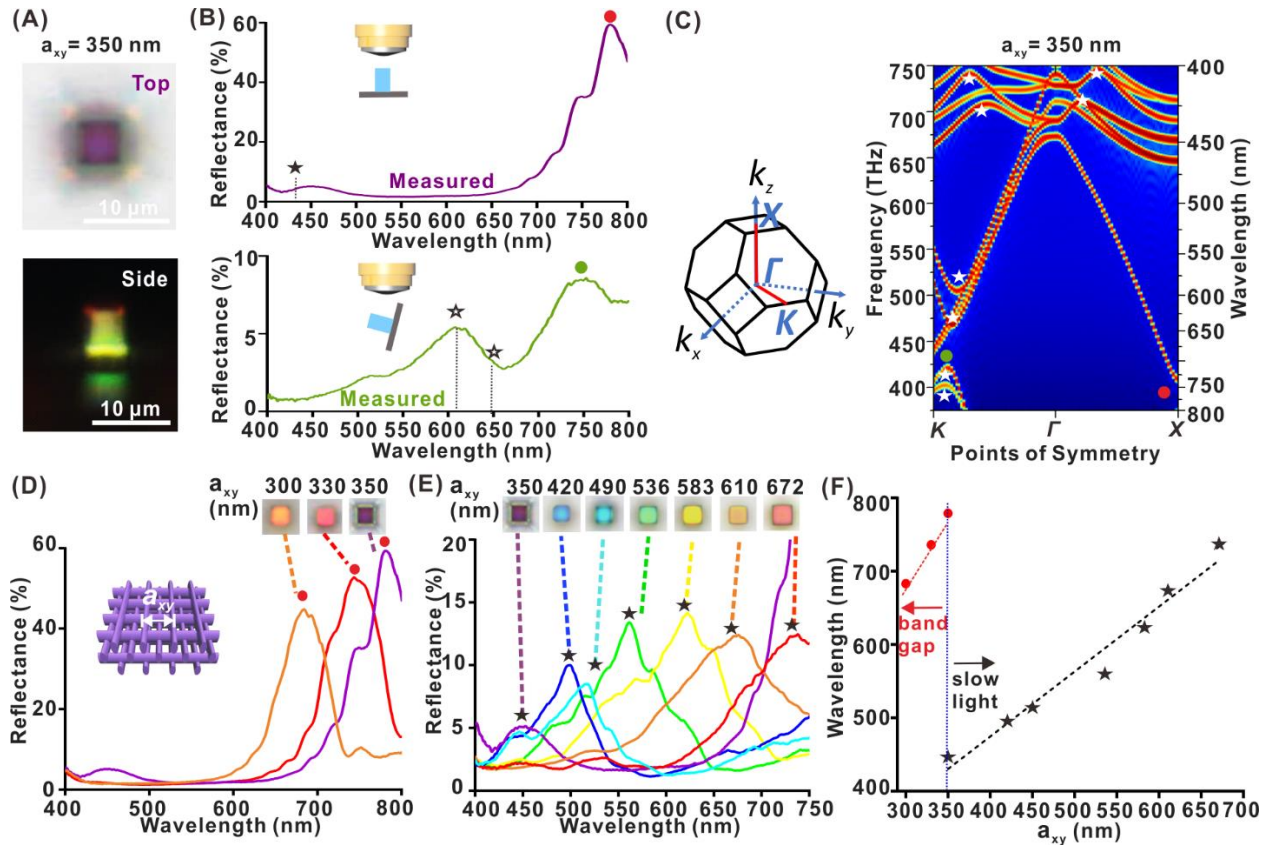

**Figure 2.** Reflectance and bandstructure of a woodpile photonic crystal with $a_{xy}$ = 350 nm, $a_z$ = 614 nm. (A) Top view (top) and side view (bottom) reflection-mode optical micrographs of the woodpile photonic crystal. (B) Reflectance of the woodpile photonic crystal measured with top-down illumination (top) and side illumination (bottom). (C) First Brillouin zone and photonic bandstructure of the woodpile photonic crystal in the Γ-K and Γ-X directions. Stars indicate slow light modes and dots indicate stopbands. (D,E) Reflectance spectra and reflection-mode micrographs of woodpiles under top-down illumination conditions for $a_{xy}$ = 300-350 nm (D) and $a_{xy}$ = 350-672 nm (E), respectively. (F) Plot of reflectance-peak positions as a function of the lattice constant.

The reflectances of woodpile photonic crystals with varying $a_{xy}$ under normal illumination are plotted in Figures 2D and 2E. For small $a_{xy}$ < 350 nm, strong peaks arise from stopbands in the NIR range and gradually blue-shift into the visible spectrum as $a_{xy}$ was decreased to 300 nm

(Figure 2D). For $a_{xy}$ > 350 nm, the main peaks in the visible originate from slow light modes (Figures 2E and S9). A systematic red-shift of the reflectance peaks is observed as $a_{xy}$ increases from 350 to 672 nm (Figure 2F). With reflection from the slow light mode being the main determinant of the color, angle-dependent colors are observed (Figures S10 and S11) as slow light modes can occur at significantly different wavelengths depending on the illumination direction. The good quantitative agreement between the experimentally observed reflection peaks and the calculated bandstructure (no fitting parameters) show that stronger reflection peaks > 700 nm originate from stopbands and weaker reflection peaks arise from slow light modes. While it is challenging to achieve stopbands below 700 nm using relatively low index polymers, the colors of the woodpile photonic crystals can still be tuned throughout the visible wavelength region by exploiting the reflection peak from the slow light mode.

The combination of the slow light mode in the visible and stopbands in the NIR gives rise to interesting possibilities such as tuning of the NIR reflectance peak while maintaining the same color in the visible. We fabricated woodpile photonic crystals with constant $a_{xy}$ = 450 nm but scaled the unit cell in the z-direction by introducing a factor A in $a_z = A\sqrt{2}a_{xy}$, as shown in Figure 3A. The woodpile photonic crystals in this series all appear blueish in the microscope images, but the reflectance spectra in Fig. 3B reveal an additional peak in the NIR that shifts significantly from ~ 800 nm to 1000 nm with increasing A. These observations agree well with the bandstructures for these photonic crystals, shown in Fig. 3C. With increasing A, the first Brillouin zone in the Γ-X direction becomes smaller and band folding occurs at smaller values of k, leading to the redshift of the stopband. Due to the slow variation of the slow light mode as a function of k, the bandstructure for shorter wavelengths remains relatively unchanged with varying A, resulting in a blue hue that depends weakly on *A*. This property can be useful in encoding information in the NIR into these structures, while maintaining a constant appearance in the visible spectrum.

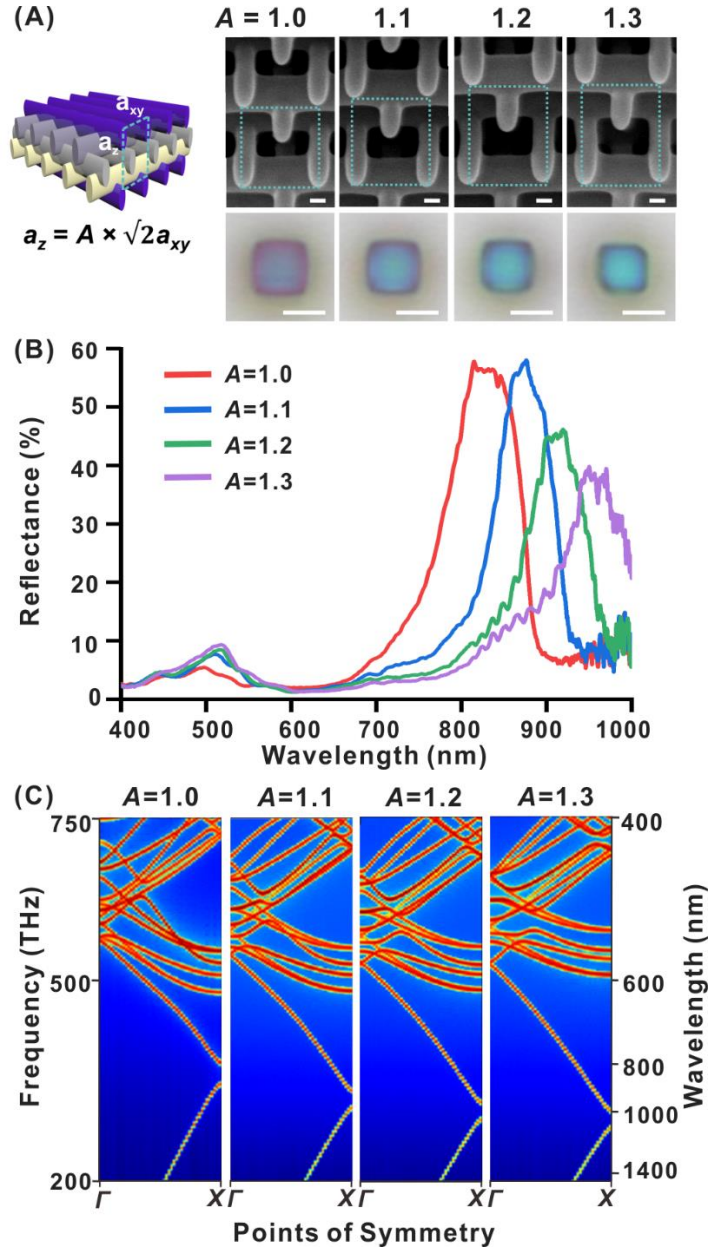

**Figure 3.** Reflectances and bandstructures of woodpile structures with fixed $a_{xy}$ = 450 nm and varying *A*, the scaling factor of $a_z$. (A) 45°-tilted-view SEM and reflection mode micrographs of the woodpile photonic crystals with *A* varying from 1.0 to 1.3. (B) Top-down reflectance spectra of the woodpile photonic crystals. (C) Bandstructures in the *Γ-X* direction for *A* = 1.0-1.3.

**3D Structural Color Printing**

To demonstrate the printing of 3D objects consisting of structural colors at the microscale, we printed a range of woodpile structures with varying laser powers (16 – 27 mW) and $a_z = \sqrt{2}a_{xy}$, with $a_{xy}$ ranging from 1.1 to 2.9 µm, as shown in the composite images in Figures 4A and S12. As printed, these structures show little to no color (Figure S12A). After heating, the woodpiles are reduced in size and become colorful (Figure S12B), with $a_{xy}$ ranging from 330 – 980 nm, $a_z$ from 580 –1490 nm and $w$ from 100 – 200 nm from SEM inspection. Due to complex dependence of degree of shrinkage on laser power and pattern density, the columns in the composite image no longer have the same $a_{xy}$ but can be grouped within similar filling factors instead (see Figure S12-14 for details). Structures occupying the upper right side of the composite image in Figures 4A and S12B have the largest $a_z > 1000$ nm, thus their main reflection peaks are in the NIR region, resulting in the poorly defined colors of the woodpile structures under normal viewing. However, when viewed from the side, these structures still appear colorful due to the smaller value of $a_{xy}$, ~ 700 nm  (Figure 4A).

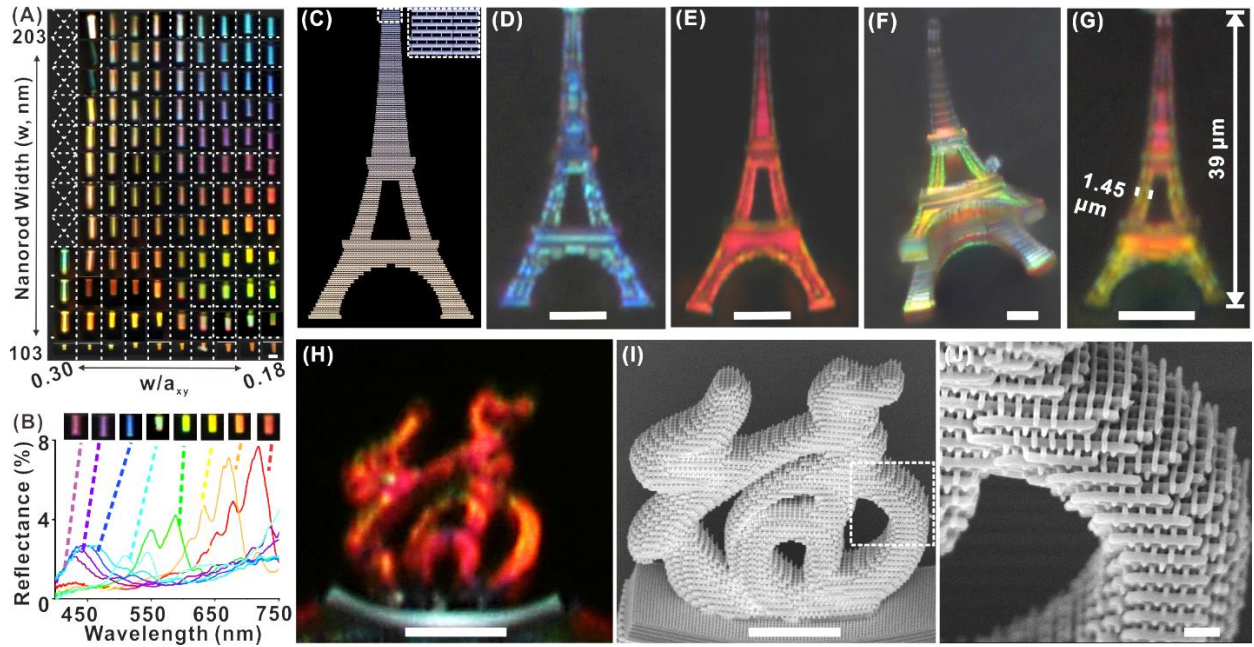

**Figure 4. 3D Color Prints.** (A) Composite optical micrographs of heat-treated woodpile photonic crystals with varying structural dimensions as viewed from the side. (B) Side illumination reflectance spectra of selected woodpile structures from (A). (C) GWL file used for lithographic printing of the *Eiffel Tower*, comprising of woodpile voxels. Micrographs of 3D printed model of the *Eiffel Tower* in structural blue (D) and structural red (E). (F) Oblique view of an *Eiffel Tower* printed with intentional gradient of colors. (G) Further down-scaled multi-color 3D print of the *Eiffel Tower*. (H) Optical micrograph and (I) SEM image of a 3D Chinese character "福" in structural red. (J) Close-up SEM image of dotted square region in (I). Scale bars in (A-I) represent 10 μm and scale bar in (J) represents 1 μm.

Woodpiles with different lattice constants can be freely positioned and concatenated into a single object to achieve a 3D structural color print (Figure S14). To demonstrate the ability to print arbitrary and complex 3D color objects at the microscale level, we fabricated microscopic models of *Eiffel Towers* comprised of woodpile voxels (Figure 4). The general writing language (GWL) format layout files for the Nanoscribe were generated by filling a stereolithographic (STL) 3D model of the *Eiffel Tower* with woodpile structures with either constant or varying periodicities.

The lattice constants were chosen to produce the desired colors after shrinking (Figure 4C). The tower was attached to the substrate at the tip and the fabricated 3D structures are observed from the side with an optical microscope. Optical micrographs in Figure 4 show that the *Eiffel Towers* have robust shapes and structures, remaining intact after thermal shrinkage, and also exhibit vivid colors. A 54 µm tall *Eiffel Tower* can be 3D printed either entirely in structural blue (Figure 4D, comprised of woodpile structures with $a_{xy}$ = 380 nm and $a_z$ = 610 nm) or structural red (Figure 4E, comprised of woodpile structures with $a_{xy}$ = 470 nm and $a_z$ = 890 nm), demonstrating the wide color range and versatility of our method. The woodpile structures are structurally stable and can be used as building blocks for a variety of models. To demonstrate the versatility of the method, a 20 µm tall Chinese character for luck "福" was printed in structural red (Figure 4H, $a_{xy}$ = 470 nm, $a_z$ = 890 nm). Multi-colored objects can also be printed. We fabricated full-color 3D prints of the *Eiffel Tower* (Figure 4F) and the *ArtScience Museum* in Singapore (Figure S16). The fabricated *Eiffel tower* 3D print had a height of 39 µm and is comprised of green, orange and fuchsia color voxels (Figures 4G). As a gauge of the color printing resolution of the woodpile structures, the smallest achievable color voxel size is 1.45 µm in the *xy*-directions and 2.63 µm in the *z*-direction (Figure S16).

The heat-induced shrinking method enables one to readily exceed the resolution limit of a 3D DLW system to print 3D objects that exhibit colors due to the underlying photonic bandstructures of the constituent lattices. The good agreement between photonic bandstructure calculations and experimental results with no fitting parameters allows us to clearly identify slow light modes and stopbands as the source of spectral peaks, giving rise to a full range of colors. While we have demonstrated that this process allows one to reproducibly create uniform or patterns of colors on a single object in cross-linked resist, this process is likely extendable to inorganic resists with higher refractive indices such as TiO$_2$ and hierarchical structures that produce angle-independent colors. Our work demonstrates the ability to produce structural color within complex 3D objects at will, and could be extended to developments in compact optical components and integrated 3D photonic circuitry that operate in the visible to NIR wavelengths.

**Methods**

**Materials.** IP-Dip photoresist with a refractive index $n \approx 1.57$ (Nanoscribe Inc, Germany) was used as a negative photoresist for two-photon lithography in dip-in laser lithography (DiLL) configuration. Propylene glycol monomethyl ether acetate, isopropyl alcohol, and nonafluorobutyl methyl ether were purchased from Sigma-Aldrich. All chemicals were used without further purification. Glass slides (fused silica, 25 mm squares with a thickness of 0.7 mm) were purchased from Nanoscribe GmbH and used without further surface modification.

**Fabrication of polymeric photonic crystal structures on glass slides.** Polymeric nano and/or microstructures were fabricated using a direct laser writing system (Nanoscribe Inc., Germany). In a typical experiment, a droplet of IP-Dip photoresist was placed onto the bottom of a glass substrate and a microscope objective was raised into this liquid. This Dip-in Laser Lithography (DiLL) configuration was performed using an inverted microscope with an oil immersion lens (63×, NA 1.4) and a computer-controlled galvo stage. Pre-defined pattern files determine the positions of the laser spot and thus the shapes of the polymerized structures. A femtosecond pulsed laser centered at 780 nm wavelength with an average power of ~ 16-27 mW and a galvo-scanned writing speed of 15 mm/s were used to crosslink the resist. Unexposed photoresist was removed via immersion in propylene glycol monomethyl ether acetate for 10 min, followed by immersion in isopropyl alcohol for 5 min and nonafluorobutyl methyl ether (NFBME) for another 5 min. Finally, the samples were removed and left to dry under ambient condition.

**Thermal treatment of polymeric photonic crystal structures.** The polymeric photonic crystals were heated by using a temperature-controlled heating stage (Linkam Scientific Instruments Ltd). The sample was put in the closed chamber of the heating stage with Ar flow. The temperature in the chamber was heated from 26 °C to 450 °C at a ramp rate of 10 °C/min and was maintained at 450 °C for 17 min. After the stage heating process was stopped, the chamber was cooled down to room temperature by a water chiller.

**Bandstructure calculation.** The bandstructures were calculated using the Lumerical Finite-Difference Time-Domain software. IP-Dip was modeled as a lossless dielectric with $n = 1.57$ and $n = 1.75$ before and after heat shrinking, respectively. Electric dipoles with random orientations were placed randomly within the unit cell of the woodpile structure. Bloch boundary conditions were used, with $k_x$, $k_y$ and $k_z$ values spanning the first Brillouin zone. The electric field oscillation as a function of time was recorded, and apodized to filter out initial transient oscillations so only the oscillations belonging to the modes of the structure that propagate indefinitely remains. The fast Fourier transform of the apodized data produces the frequencies of the modes. A series of simulations with the desired ($k_x$, $k_y$, $k_z$) values were performed to produce the bandstructures. For woodpile structures where $a_z \neq \sqrt{2}a_{xy}$, the body-centered cubic (BCC) Brillouin zone is typically used. However, for ease of comparison, we used a stretched face-centered cubic (FCC) Brillouin zone and used the same symmetry points as the FCC Brillouin zone.

**Reflectance measurement.** Optical micrographs and spectra (reflectance mode) were taken using a Nikon Eclipse LV100ND optical microscope equipped with a CRAIC 508 PV microspectrophotometer and a Nikon DS-Ri2 camera. Samples were illuminated with a halogen lamp and measured/imaged in reflection mode though a 50×/0.4 NA long working distance objective lens. The angle-varying reflectance measurements were performed on a tilt stage. The tilt angle of the stage can be tuned from 0° to 90° with 15° steps. The incident and reflected light paths were normal to the substrate when the tilt angle is 0°. The spectra (reflectance mode) are normalized to the reflectance spectrum of aluminum, which is measured under the same conditions as for the sample.

**Characterization.** Scanning electron microscopy (SEM) was performed using a JEOL-JSM-7600F SEM system with an accelerating voltage of 5 kV. Thermogravimetric analysis (TGA) was performed using a thermogravimetric analyzer, TA Q50, with a Pt boat to hold the sample. The entire TGA analysis was performed in a closed chamber with $N_2$ flow (60 ml/min).

**Acknowledgements**

We acknowledge funding support from the National Research Foundation grant award No. NRF-CRP001-021, A*STAR Young Investigatorship (Grant 0926030138), SERC (Grant 092154099) and SUTD Digital Manufacturing and Design (DManD) Center grant RGDM1830303. We thank Robert Edward Simpson, Weiling Dong and Tian Li for technical support with the temperature-controlled heating stage.

# Supplementary Figures for Structural Color 3D Printing By Shrinking Photonic Crystals

**Authors:** *Yejing Liu,[1] Hao Wang,[1] Jinfa Ho,[3] Ryan C. Ng,[2] Ray J. H. Ng,[1,3] Valerian H. Hall-Chen,[4] Eleen H. H. Koay,[3] Zhaogang Dong,[3] Hailong Liu,[1] Cheng-Wei Qiu,[5] Julia R. Greer,[2] Joel K. W. Yang,[1,3,*]*

[1] Engineering Product Development, Singapore University of Technology and Design, Singapore 487372.

[2] Division of Engineering and Applied Science, California Institute of Technology, Pasadena CA 91125, United States of America.

[3] Nanofabrication Department, Institute of Materials Research and Engineering, Singapore 138634.

[4] Rudolf Peierls Centre for Theoretical Physics, University of Oxford, Oxford OX1 3PU, United Kingdom.

[5] Department of Electrical and Computer Engineering, National University of Singapore, Singapore 117583.

* To whom correspondence should be addressed. Email: joel_yang@sutd.edu.sg

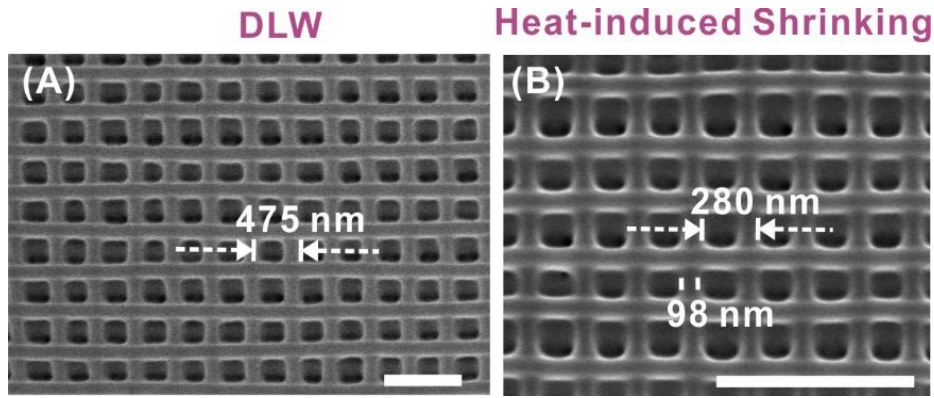

**Figure S1.** Woodpile separate structures at the resolution limits of the conventional DLW and heat-induced shrinking method. SEM images of (A) a woodpile with $a_{xy}$ = 475 nm fabricated with DLW without heating and (B) a different woodpile with $a_{xy}$ = 280 nm fabricated with the heat-induced shrinking method. The lattice constant of this woodpile prior to shrinking was 700 nm. Scale bars represent 1 μm. Parameters for patterning structure in (A): Write speed = 1 mm/s, laser power = 7.5 mW, nominal pitch 600 nm (note some shrinkage occurs after sample development). Parameters for patterning structure in (B): Write speed = 15 mm/s, laser power = 14.5 mW, nominal pitch preset = 700 nm.

**Advantages and additional characterization of heat shrinking process**

Figure S2A and S2B compares woodpile structures fabricated by conventional direct laser writing (DLW) and with heat induced shrinking. A key advantage of the heat shrinking process is the widening of the process window, enabling the fabrication of structures with smaller periods than is possible with DLW. Figures S2C and S2D compare the volume fraction $\Phi$ of conventional DLW vs heat induced shrinking. $\Phi$ is calculated by:

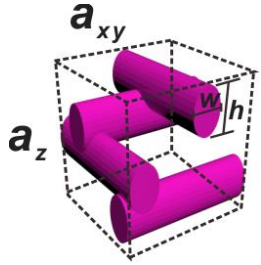

$$\Phi = 4\pi \left(\frac{w}{2}\right)\left(\frac{h}{2}\right) a_{xy} / a_{xy}^2 a_z$$

$w$: width of nanorod

$h$: height of nanorod

$a_{xy}$: lateral lattice constant

$a_z$: axial lattice constant

A $\Phi$ of 1 indicates unsuccessful fabrication where the structures have fused into a solid block. Our calculation ignores the effect of rod overlap and therefore gives a conservative estimate of the laser power and lattice constant where the structures fuse into a solid block. Woodpile photonic crystals with $a_{xy}$ = 300 to 500 nm, $a_z$ = 425 to 710 nm were fabricated with different laser powers. The smallest $w$ and $h$ of the constituent rods were obtained from SEM images, and $\Phi$ was calculated using a unit cell containing four rods with the arrangement shown in Figure S2C. With conventional DLW, < 10% of the structures fabricated yielded $\Phi$ < 1 with parameters $a_{xy}$ > 475 nm and laser power < 17.5 mW. The smallest $a_{xy}$ successfully fabricated was 475 nm with $\Phi$ = 0.9 (Figure S1A). Attempts at fabricating woodpiles with smaller lattice constants by using low laser power (7.5 mW) resulted in collapsed structures (Figure S3C). With heat shrinking, well-defined and structurally sound woodpile photonic crystals with $a_{xy}$ of 300 to 500 nm were successfully fabricated. Well-defined woodpile structures with $a_{xy}$ ~280 nm and rod width ~100 nm were achieved (Figures S1B).

In addition to decreasing the size and period of the 3D printed structure, heat shrinking also alters the effective shape of the writing spot, making it more spherical. As the Gaussian beam profile at the point of exposure has a depth of focus that is ~ 3x larger than the beam waist, the rod structures exposed with a single laser pass have elliptical cross-sections, as shown in Figure S2A. After heat-induced shrinking, the cross-sections become more circular, as shown in Figure S2B. The aspect ratio $h/w$ for pre-heat shrunk structures is ~ 2.5 – 3.4, depending on the exposure power of the laser (Figure S5). During the heat-induced shrinking process, there is a larger decrease in height (39 – 60%) compared to the decrease in width (30 – 53%). The difference in shrinking is most pronounced in IP-Dip structures that have low degrees of cross-linking (exposed at low laser powers). With a laser power of 16 mW, the rods have $h/w$ = ~210 nm / 105 nm = 2 after heat shrinking. The tendency of the structures to become more spherical suggests surface energy minimization during the shrinking process.

Other than woodpile photonic crystals, heat induced shrinking can also be used to fabricate other photonic crystal designs. To demonstrate this, we successfully fabricated 3D photonic crystals with gyroid[1] and diamond lattices[2,3] with periods of 375 and 290 nm, respectively (Figures S2E and S2F). Such small feature sizes are comparable with previous reports of structures fabricated using STED.[4] Crucially, these structures could be achieved with higher throughput and/or lower laser powers as there is no competition between excitation and de-excitation processes here. Typically, it merely takes less than 30 s for printing a woodpile structure with dimension of 20 μm × 20 μm × 32 μm ($a_{xy}$=1.5 μm, $a_z$=2.1 μm).

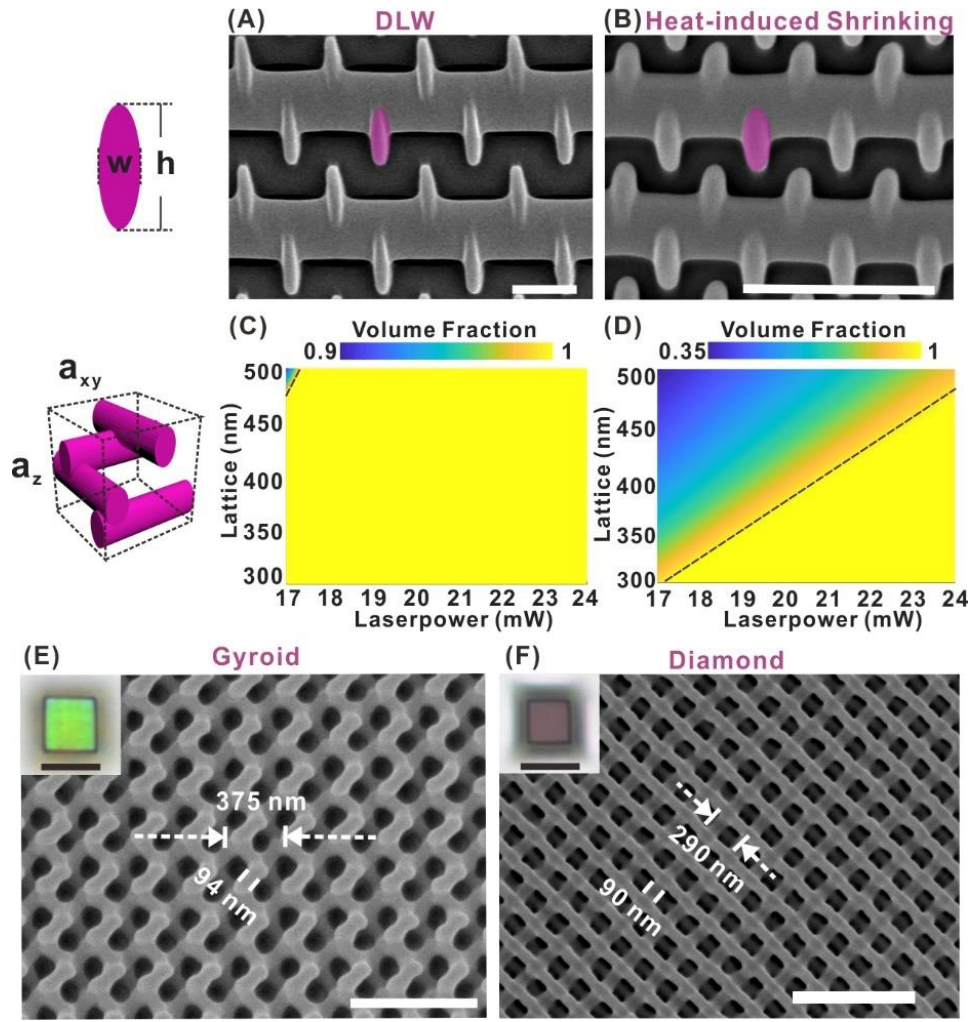

**Figure S2. The effect of heat shrinking on the effective writing spot shape, process window and the successful patterning of different photonic crystals.** Side-view SEM images of woodpile photonic crystals fabricated by (A) direct-laser-writing with laser power of 20 mW and writing speed of 15 mm/s and (B) after heat treatment. Volume fraction maps of unit cells with lattice sizes less than 500 nm and made from various lithography laser powers, for (C) direct-laser-writing and (D) our proposed heat treatment method, respectively. Optical micrographs (insets, scale bars represent 10 μm) and SEM images of (E) a gyroid photonic structure and (F) a diamond photonic crystal fabricated using the heat-induced shrinking method. Scale bars represent 1 μm.

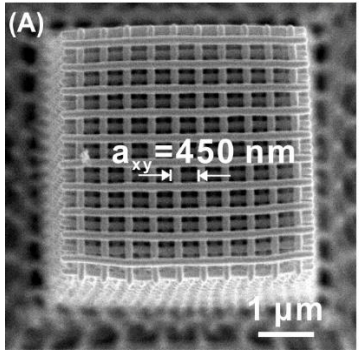 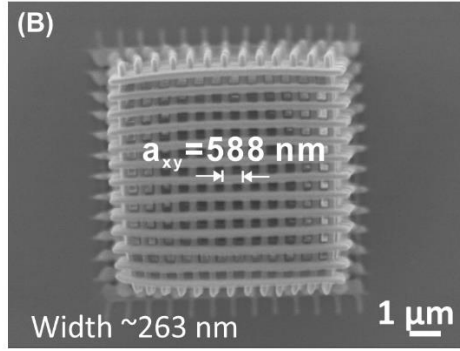 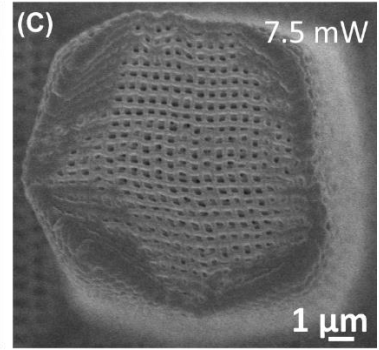

**Figure S3.** Comparison of woodpile structures fabricated using conventional direct-laser-writing (DLW) and with the heat-induced shrinking method. (A) With heat treatment, we were able to fabricate a structurally well-defined woodpile structure with $a_{xy}$ = 450 nm (nearly the resolution limit of our Nanoscribe system) with well separated neighboring rods. (B, C) SEM images of woodpile structures with $a_{xy}$ = 588 nm fabricated with conventional DLW. The rods were connected at various regions (B). This cannot be solved by adjusting exposure parameters, e.g. by lowering the laser power or write speed. Below the polymerization threshold of 7.5 mW, the IP-Dip was not sufficiently crosslinked and does not survive the development step, resulting in collapsed woodpile structures (C).

**Estimated axial resolution (resolution in z-direction)**

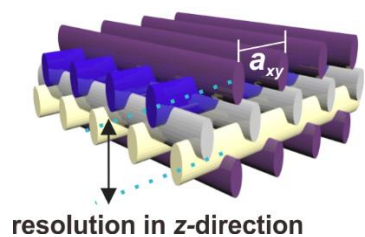

**Figure S4.** Scheme showing axial resolution ($R_z$) in a woodpile structure. The axial resolution is described by the smallest center-to-center distance between two separated nanorods in z-direction, which is $3a_z/4$.

For lithographic woodpile structure before heating:

$a_z = \sqrt{2}a_{xy}$ (designed in lithography)

$R_z = 3\sqrt{2}a_{xy}/4$

When $a_{xy} = 475$ nm

$R_z = 475 \times 3\sqrt{2}/4 = 504$ nm

After heat-induced shrinking:

$a_z = S \times \sqrt{2}a_{xy}$, S is the factor of anisotropic shrinking in $xy$- and $z$-direction.

$S = 1.89$

$R_z = 1.89 \times 3\sqrt{2}a_{xy}/4$

When $a_{xy} = 280$ nm

$R_z = 380$ nm.

It is noteworthy that a good axial resolution is more challenging to achieve than a lateral resolution for DLW, as the axial Abbe limit is 2.92 times larger than the lateral Abbe limit.[4] With $a_{xy} = 475$ nm (smallest period that can be fabricated with our Nanoscribe system), the resolution limit in $z$-direction is estimated to be 504 nm (see calculation above). Using the heat-induced shrinking method, the smallest $a_{xy} = 280$ nm, corresponding to a resolution limit of 380 nm in the $z$-direction, which is below the two-photon Sparrow criterion in $z$-direction of 506 nm and comparable with the ultimate values achieved with STED ($a_{xy} = 275$ nm and $z$-resolution 375 nm).[4]

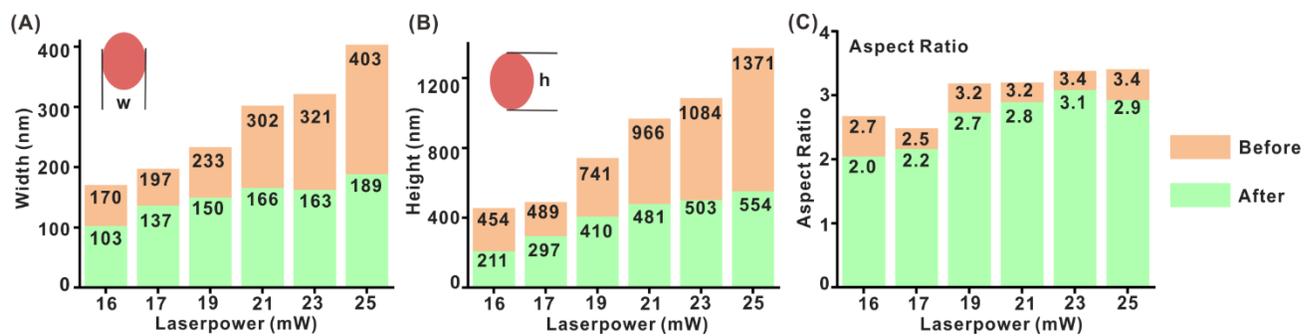

**Figure S5.** Stack-bar charts showing dimensions of the effective writing spot before (orange) and after (green) heating for (A) the rod width, (B) the rod height and (C) the aspect ratio.

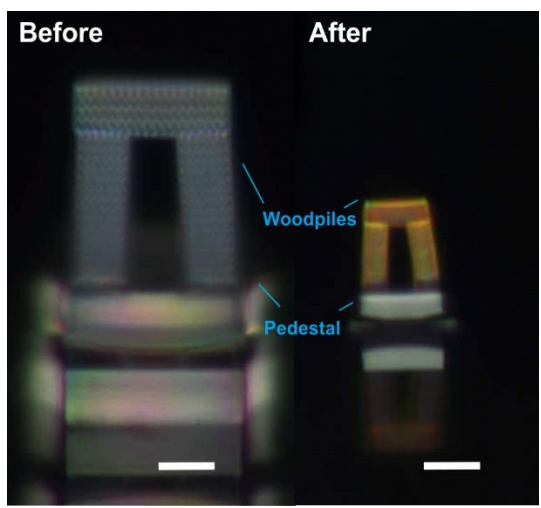

**Figure S6.** Side-view images of photonic crystal structures sitting on pedestals made of fully cross-linked IP-Dip before and after thermal shrinking. All scale bars represent 10 µm.

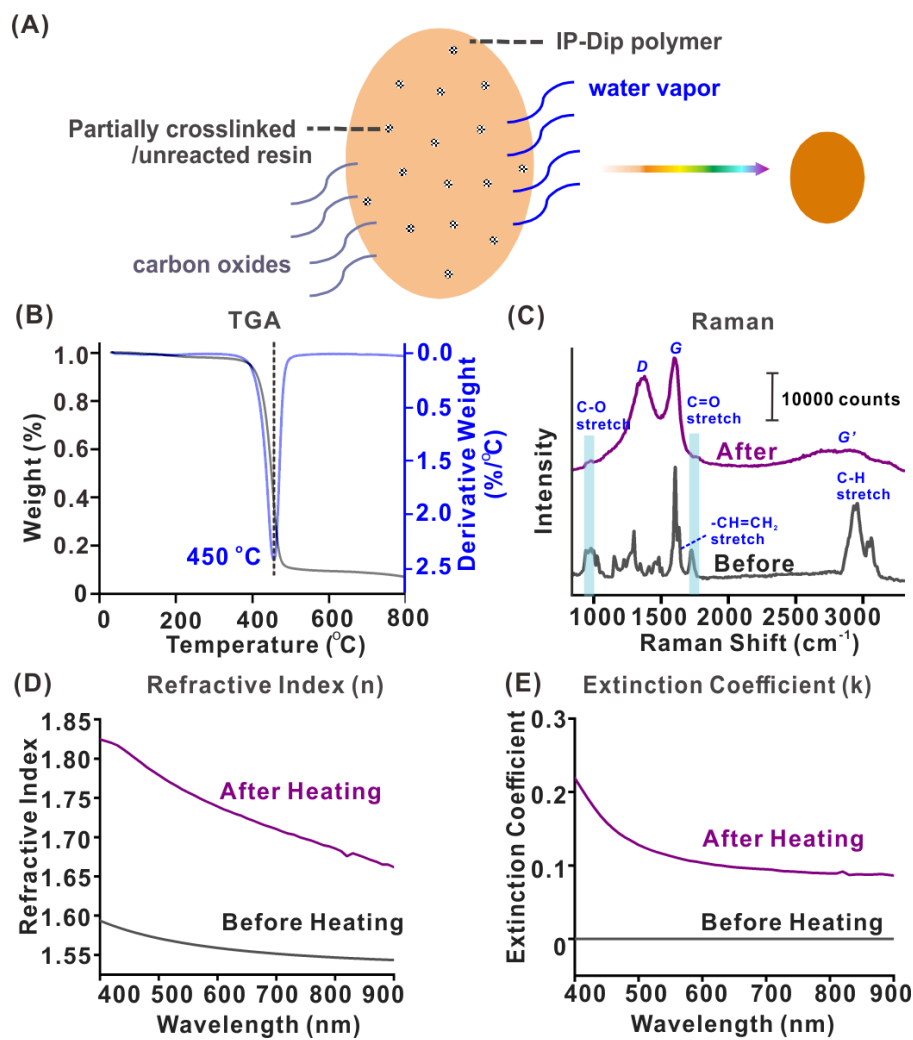

**Figure S7.** Characterization of IP-Dip before and after thermal treatment. The main components of IP-Dip are 2-(hydroxymethyl)-2-[[(1-oxoallyl)oxy]methyl]-1,3-propanediyl diacrylate (CAS: 3524-68-3, 60-80%), 9H-fluorene-9,9-diylbis(4,1-phenyleneoxyethane-2,1-diyl)-bisacrylate (CAS: 161182-73-6, < 24%), and Biphenyl-2-ol, ethoxylated, esters with acrylic acid (CAS: 72009-86-0, < 24%) (A) Schematic showing the process by which the IP-Dip polymer loses volume, resulting in increased density and more carbon content. During the heating process, partially crosslinked resin (or unreacted resin) within the cured IP-Dip was removed. Pyrolysis of the polymer causes carbon oxides and water vapor to be released. (B) Thermogravimetric analysis of IP-Dip heated from 0 to 800 °C. (C) Raman spectra of IP-Dip before and after heating. (D) Refractive indices and (E) extinction coefficients of IP-Dip before and after heating.

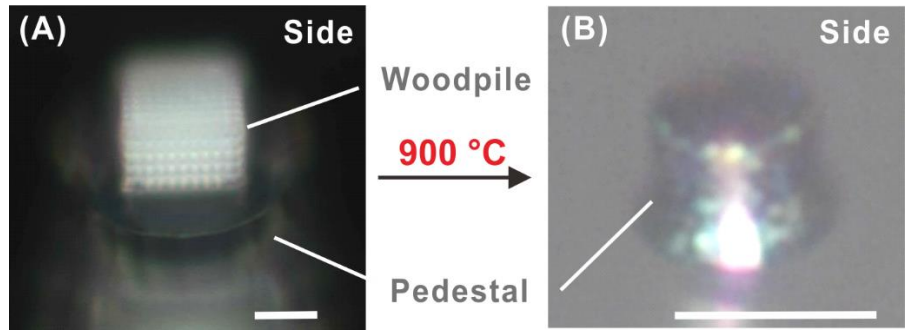

**Figure S8.** Woodpile structure ($a_{xy}$ = 1.9 µm, $w$ = 403 nm) under pyrolysis at 900 °C. Side-view reflection images of the woodpile sitting on the pedestal made of fully cross-linked IP-Dip (A) before, and (B) after pyrolyzing at 900 °C. Woodpile did not survive in the pyrolysis and only the solid pedestal which is more structurally stable remained intact after heating to 900 °C. Scale bars represent 10 µm.

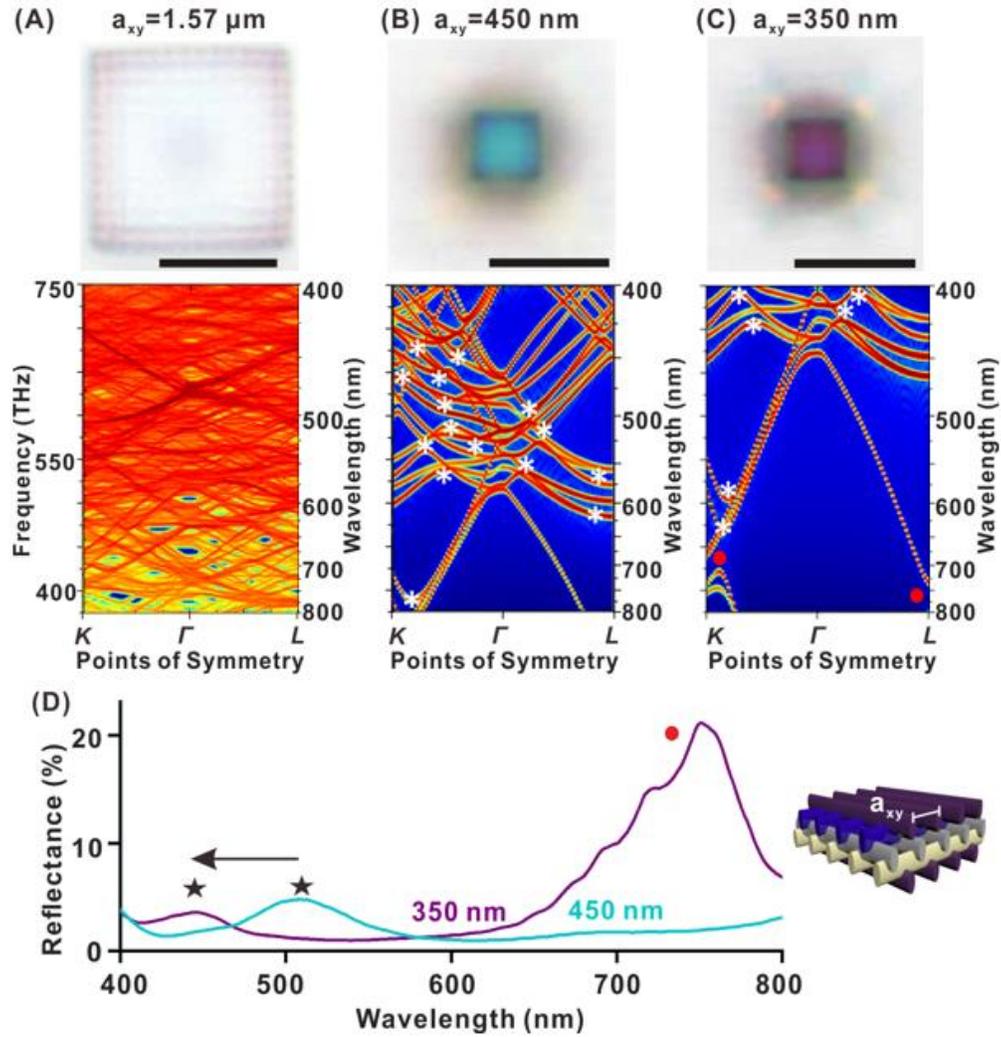

**Figure S9.** Micrographs (top) and band structure diagrams (bottom) of woodpiles with (A) $a_{xy}$ =1.57 µm before heating, and (B) $a_{xy}$ = 450 nm and (C) $a_{xy}$ = 350 nm. (D) Reflectance spectra for woodpiles with $a_{xy}$ = 450 nm (blue) and $a_{xy}$ = 350 nm (purple).

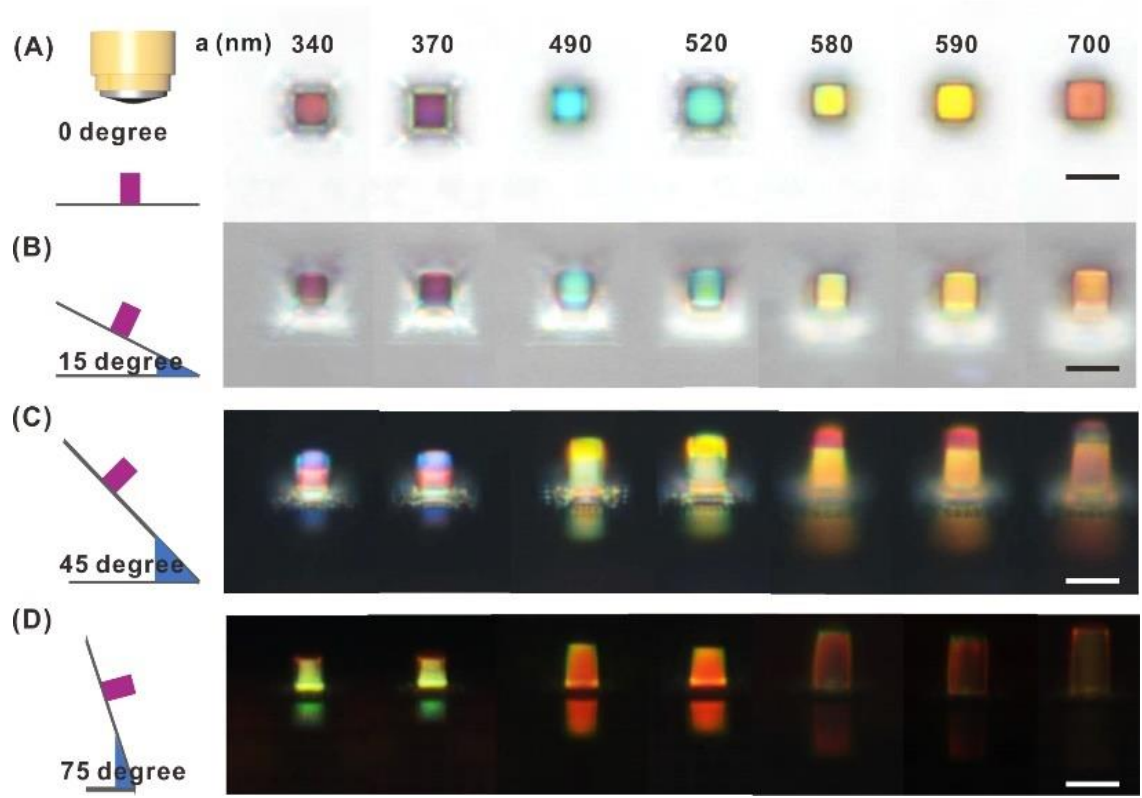

**Figure S10.** Refection-mode optical microscope images showing angle-dependent color from woodpile structures with periods of 340-700 nm. The micrographs are taken at tilt angles of (A) 0º, (B) 15º, (C) 45º and (D) 75º. All scale bars represent 10 μm.

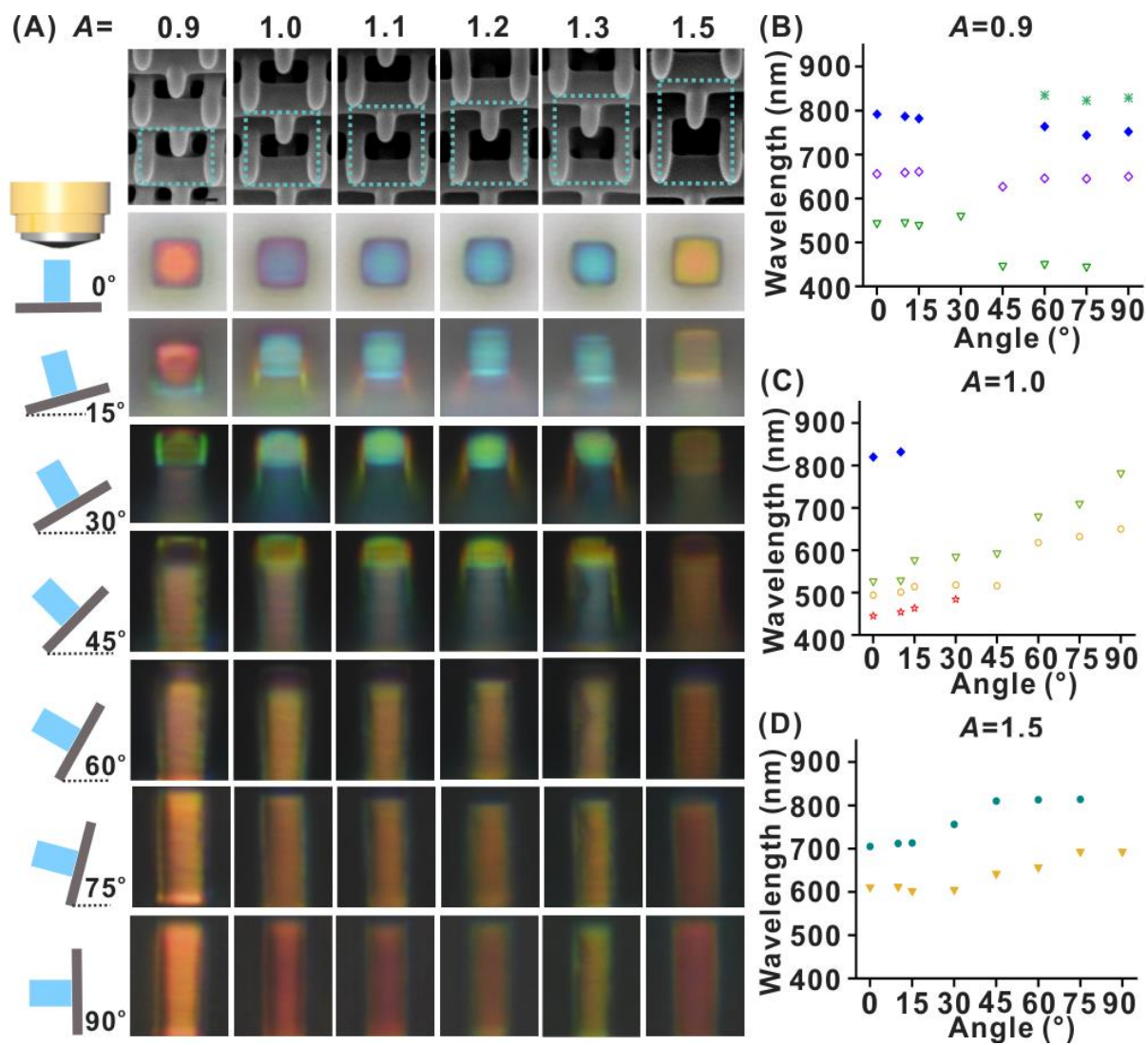

**Figure S11.** (A) SEM (top panel) and reflection-mode micrographs of woodpiles fabricated with lateral period ($a_{xy}$) of ~450 nm and scaling factor A varying from 0.9 to 1.5, viewed at tilt angles of 0° to 90°. Plot of the peak wavelengths in the reflectance spectra measured at tilt angles 0° to 90° for (B) *A*=0.9, (C) *A*=1.0 and (D) *A*=1.5, respectively.

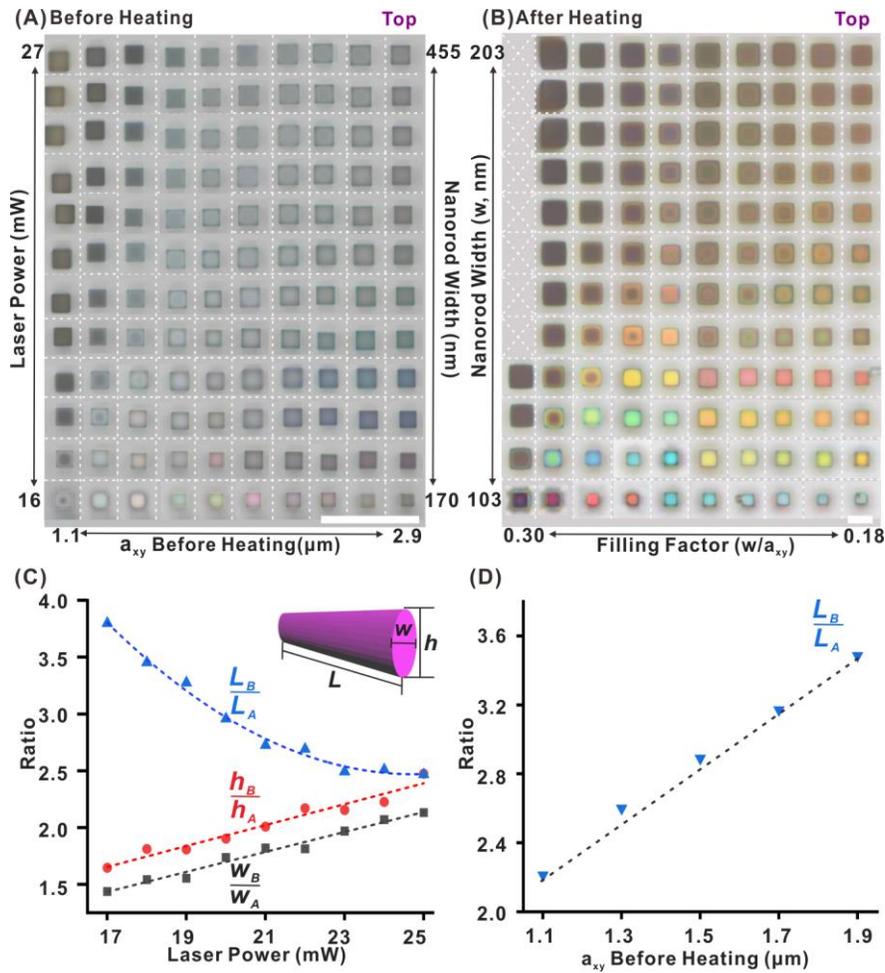

**Figure S12.** 3D color voxels exhibiting a wide range of colors. (A) Top view reflection-mode micrographs of woodpile photonic crystals before heat shrinkage. $a_{xy}$ varies from 1.1 to 2.9 µm across each row (step value 0.2 µm), and the laser power used during the writing process varies from 16 to 27 mW across each column (step value 1 mW). The rod width is a function of the laser power, and varies from 170 to 455 nm. (B)Top-view reflection-mode micrographs of the woodpile photonic crystals after heating. Heating the dense woodpiles produced with laser power from 20-27 mW induces their change into solid blocks and these solid structures are removed from B (in the first column). The lattice constant increases diagonally with colors shifting from magenta to blue, green, yellow, orange, and finally red. (C) Plot of $L_B/L_A$, $h_B/h_A$, and $w_B/w_A$ for rods in woodpiles exposed using laser powers of 17-25 mW, where $L_B$ ($L_A$), $h_B$ ($h_A$), and $w_B$ ($w_A$) denote the length, height and width of the rods before (after) heating, respectively. The dotted lines are fitting curves. (D) Plot of $L_B/L_A$ for rods fabricated with direct laser writing (without heat

shrinkage), with $a_{xy}$ varying from 1.1 to 1.9 μm. The scale bars in A and B represent 100 μm and 10 μm, respectively.

**Different Shrinkage Rates**

The manner in which the woodpile structures shrink was investigated and plotted in Fig. S13C and S13D. The shrinkage was characterized by the ratio of each dimension measured from SEM images before and after heating, i.e. $w_B/w_A$, $h_B/h_A$ and $L_B/L_A$. We observe that with increasing laser power, i.e. higher degree of crosslinking, the rod shrinkage decreased along the length of the rod but increased along the shorter axes, i.e. $h$ and $w$. Furthermore, for a given laser power, the degree of shrinkage increases with increasing period. These results show intuitively that the largest shrinkage is achieved for sparse, weakly crosslinked structures.

The heat-induced shrinkage is anisotropic in width ($w$), height ($h$) and length ($L$) of nanorods in woodpile structures. At certain exposure laser power, the shrinkage follows $w_B/w_A < h_B/h_A < L_B/L_A$ to reduce the surface energy, indicating that the lattice constant in $xy$-direction reduces more than in $z$-direction for woodpile structures (Fig. S12C). In this case, the geometry of the unit cell is elongated along z after heating and the factor of elongation ($E$) is a function of laser exposure power (Fig. S13A). In order to fabricate woodpile unit cells with a certain geometry ($a_z/a_{xy}$), a compensating factor ($C$) can be multiplied to $a_z$ in design such that $C=1/E$ (Fig. S13B).

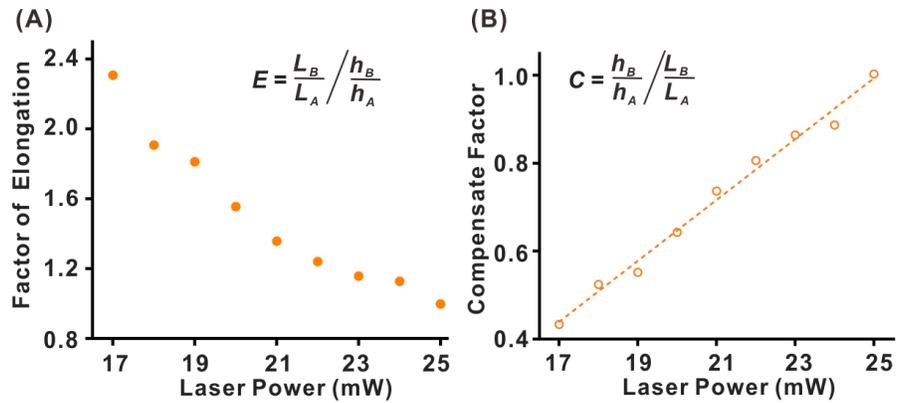

**Figure S13.** Plot of (A) factor of elongation (*E*) and (B) compensation factor for anisotropic shrinkage in *xy*- and *z*-direction, as a function of laser power.

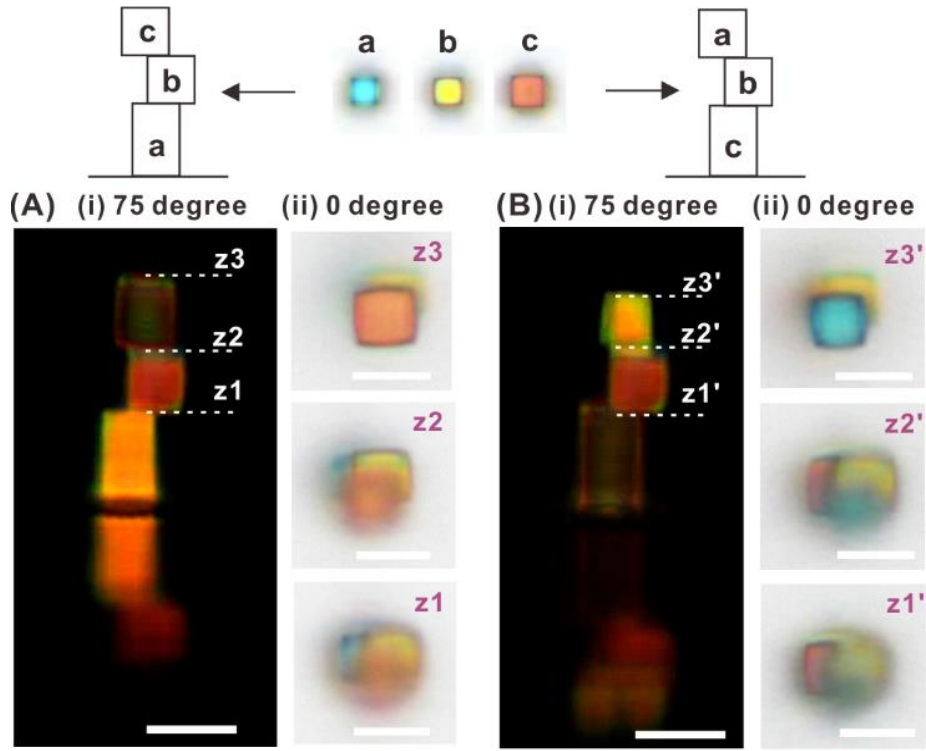

**Figure S14.** Colored "ladder" (3D stacked cubes) demonstrating a simple 3D printed color object. (A, B) Reflection-mode optical micrographs of two 3D color voxels stacks. *a*, *b* and *c* are woodpile voxels with different periods. Voxel *a* is placed at the bottom in (A) and at the top in (B). Reflection-mode optical micrographs are taken at tilt angles of (i) 75 degree and (ii) 0 degree, and optical micrographs images are taken at three focal planes in (ii). All scale bars represent 10 μm.

We constructed two "ladder" structures (3D stacked cubes) to demonstrate the ability to print 3D color objects and its multiplexing capability. These two ladders both comprise of three color voxels (a, b and c) with different stacking orders. Voxels a, b and c were chosen from the library of shrunken woodpiles exhibiting cyan, yellow and orange colors at normal incidence (0 degree). Their periods are 500, 580 and 700 nm, respectively. After thermal shrinking, the ladder structures exhibit vivid colors, with different colors appearing at different focal planes and viewing angles. Reflection-mode optical micrographs demonstrate that the colors of the three voxels could be read out independently at both viewing angles. The shrinking percentage and the resultant color were almost identical for woodpile structures with the same lattice constant in the two stacks, and the shrinking can therefore be considered to be independent of neighboring voxels. As such, we could arbitrarily put the color voxels at the top or at the bottom and individually read out them via reflectance colors observed from side, indicating the multiplexing capability of our strategy.

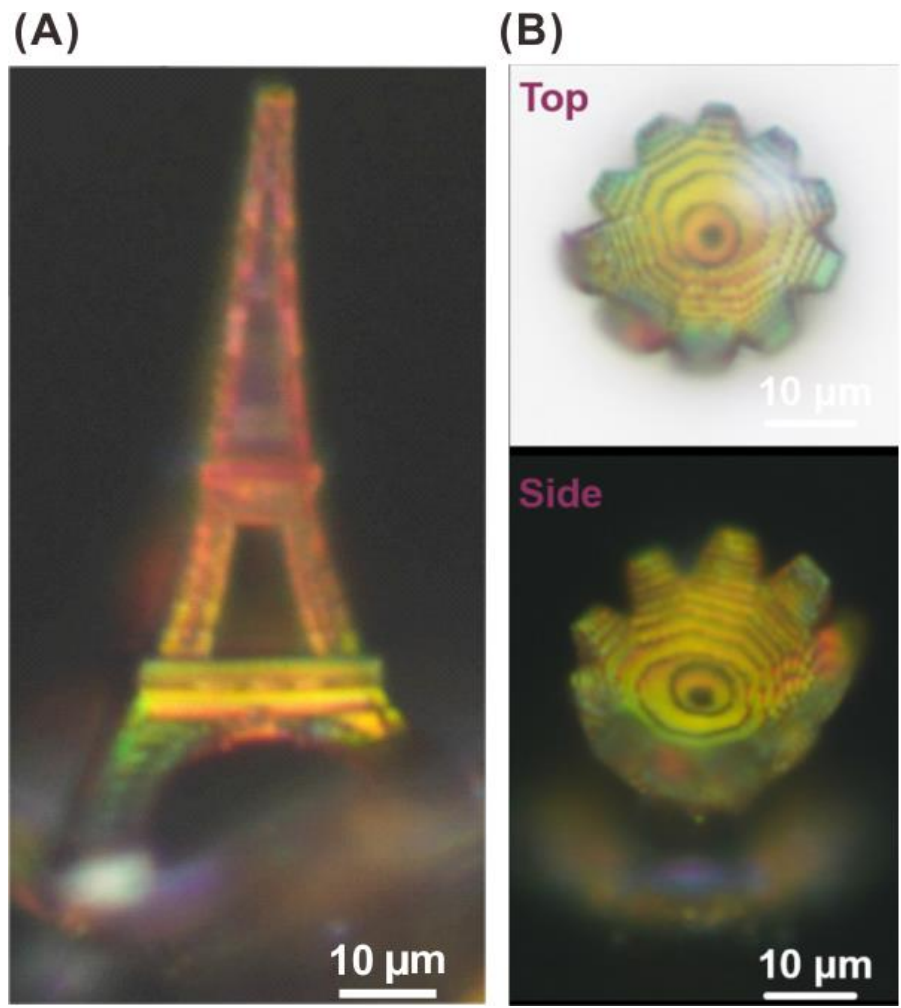

**Figure S15.** Full-color models printed with heat induced shrinking method. Optical micrographs of models of the (A) *Eiffel Tower* and (B) *ArtScience Museum* in Singapore.

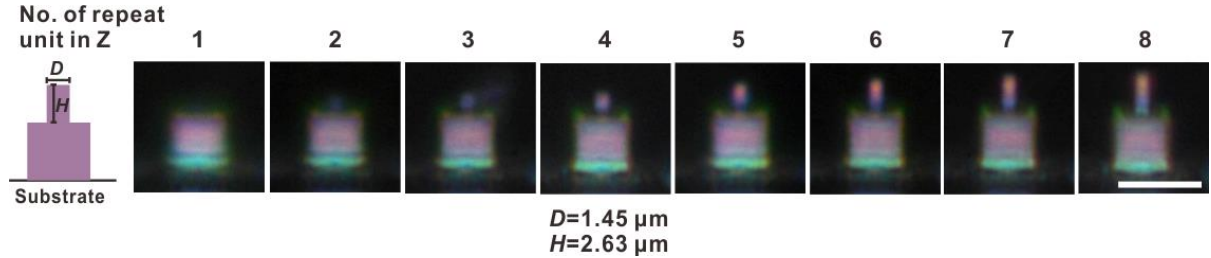

**Figure S16.** Reflection-mode micrographs for voxel height testing. On top of the sacrificial pedestal, thin pillars made of woodpile structures are fabricated at a fixed $D$ of 1.45 μm which is the smallest achievable. The height of the pillars is increased from left to right by increasing the number of repeat units in the $z$-direction, from 1 to 8. We observe from the micrographs that color can be seen only when the number of repeating units is more than 4, where the height is 2.63 μm. This represents the smallest voxel that we can fabricate. Scale bar represents 10 μm.

**Supplementary References**